\documentclass[aps,reprint,showpacs,amsmath,amssymb,prc,floatfix,nofootinbib]{revtex4-1}
\usepackage{color}
\usepackage{graphicx}
\usepackage{dcolumn}
\usepackage{bm}
\usepackage[colorlinks,citecolor=blue,linkcolor=red,anchorcolor=blue,filecolor=blue,urlcolor=blue]{hyperref}
\usepackage{times}
\usepackage{float}
\begin{document}
\title{Systematic study for the surface properties of neutron star}
\author{Ankit Kumar$^{1,2}$}
\email{ankit.k@iopb.res.in}
\author{H. C. Das$^{1,2}$}
\email{harish.d@iopb.res.in}
\author{Jeet Amrit Pattnaik$^3$}
\email{jeetamritboudh@gmail.com}
\author{S. K. Patra$^{1,2}$}
\email{patra@iopb.res.in}
\affiliation{\it $^{1}$Institute of Physics, Sachivalaya Marg, Bhubaneswar 751005, India}
\affiliation{\it $^{2}$Homi Bhabha National Institute, Training School Complex,
Anushakti Nagar, Mumbai 400094, India}
\affiliation{$^3$Department of Physics, Siksha 'O' Anusandhan, Deemed to be University, Bhubaneswar-751030, India}
\date{\today}
\begin{abstract}
\begin{description}
\item[Background] In our earlier work {\bf [Phys. Rev. C 104, 055804 (2021)]}, we studied the surface properties of a neutron star, assuming it as a huge finite nucleus containing protons, neutrons, electrons, and muons. For the first time, we reported these results of a neutron star for a few representative masses. In the present paper, we give a detailed study of these quantities to draw definite conclusions. 

\item[Method]To carry forward our earlier idea, the energy density functional of the momentum space of neutron star matter is converted to the coordinate space in a local density approximation. This functional is again used to derive the neutron star surface properties within the coherent density fluctuation model using the weight function obtained from the density profile of the neutron star using the recently developed G3 and widely used NL3 and IU-FSU parameter sets in the context of relativistic mean-field formalism.

\item[Results] The systematic surface properties of the neutron star, such as incompressibility, symmetry energy, slope parameter, and curvature coefficient, is calculated. The volume and surface components of the total symmetry energy are decomposed with the help of the $\kappa$ factor obtained from the volume to surface ratio of the symmetry energies in the liquid drop limit of Danielewicz. The magnitude of the computed surface quantities increases with the neutron star's mass. 
\item[Conclusion] The incompressibility $K^{\rm star}$, symmetry energy $S^{\rm star}$, slope parameter $L_{\rm sym}^{\rm star}$ and curvature coefficient $K_{\rm sym}^{\rm star}$ of the neutron stars with different mass are analyzed and found to be model dependent. NL3 is the stiffest equation of state endue us with the higher magnitude of surface quantities as compared to the G3 and IU-FSU forces.
\end{description}
\end{abstract}

\maketitle
\section{Introduction}
Among all the known objects in the universe, the neutron star (NS) is considered one of the densest. It is known that the cores are $10^{3}$ times or more denser than the density at the ``neutron drip" line \cite{PhysRevLett.73.2650}. To understand the properties of NS, a thorough knowledge of both nuclear physics and astrophysics is demanded. In a broad concept, the NS can be treated as a giant asymmetric nucleus comprised mostly degenerated neutrons gas with a small fraction of protons and electrons to maintain a charge-neutral body along with some exotic particles, such as hyperons \cite{bksharma07}. The NS is bounded by the attractive gravitational force balanced through the short-range strong nuclear interaction generated by the baryons and mesons. In addition to these two opposite forces, the electromagnetic interactions are also vital for NS's stability. In contrast, the normal nucleus is governed by strong nuclear and relatively weaker electromagnetic interactions. As a consequence, inside a nucleus, the density is flat, and inside the NS, the density increases. Because of this difference in the distribution of corresponding densities, it is not straightforward that the behavior of similarly defined properties is the same. One can not generalize the specific nuclear properties to the NS and needs a separate analysis to understand the NS properties.

In Ref. \cite{PhysRevC.103.024305},  the conventional Br\"uckner energy density functional  (B-EDF) \cite{bruck68,bruck69} is replaced by the effective field theory motivated relativistic mean-field (E-RMF) density functional in a local density approximation (LDA) to calculate the nuclear surface properties in the framework of the coherent density fluctuation model (CDFM). They demonstrated that the microscopic E-RMF energy functional is able to incorporate the structural effects of the nucleus and reproduce the peak in the symmetry energy of the Pb isotopic chain at neutron number $N=126$, which is generally failed by the B-EDF functional \cite{abdul20,gaid11}. Recently, this formalism has been extended successfully to study the surface properties of NS \cite{ankit2021} and estimated for the first time the incompressibility $K^{\rm star}$, symmetry energy $S^{\rm star}$, slope parameter $L_{\rm sym}^{\rm star}$ and curvature coefficient $K_{\rm sym}^{\rm star}$ for some specific masses of the star. These properties of NS are pretty informative for experimental observations and theoretical modeling. For example, assuming the NS as a giant nucleus,  with the mass number $A\sim 10^{57}$, it must possess most of the properties of a standard finite nucleus \cite{ankit2021}. It should have the multipole moments and all possible collective oscillations like $f$-mode and $g$-mode, etc. \cite{PhysRevD.104.123006}. 

The parameters obtained by expanding the symmetry energy near the saturation density (slope and curvature parameter) control the cooling rate of NS, and the core-crust transition density and transition pressure \cite{PhysRevC.94.052801, PhysRevC.100.055802}. These parameters also play an important role to constraining the nuclear equation of state (EoS), which is a key ingredient for the study of NS properties as well as the properties of supernovae explosion, binary NS merger, and the physics of gravitational wave \cite{PhysRevC.103.034330}. A prominent bridge between the finite nuclei and the nuclear/neutron matter (interstellar bodies) is the comprehensive knowledge of the nuclear EoS. The EoS is the key component for the determination of the properties of NS, and also it controls the dynamics of core-collapse supernovae remnants, and the cooling of NS \cite{PhysRevC.100.055802, Bombaci_2018}. With the help of observational gravitational wave (GW170817), \cite{PhysRevLett.119.161101}, Einstein Observatory (HEAO-2) \cite{BOGUTA1981255} and X-ray radio telescopes \cite{NASA, Greif_2020}, a large number of constraints had been implemented to get a proper EoS at high density regime. 
  
From last few decades, the non-relativistic (Skyrme \cite{doi:10.1098/rspa.1961.0018,CHABANAT1997,CHABANAT1998,DUTRA2008,DUTRA2012}, Gogny forces \cite{PhysRevC.21.1568}) and relativistic \cite{WALECKA1974491,REINHARD86,RING87,GAMBHIR1990132,MULLER96,PATRA01,PATRA02,ARUMUGAM04}) theoretical approaches have been used as consistent formalism to construct the EoS and calculate the properties of strongly-interacting dense matter systems. The relativistic class of models is the alternative approach for low-energy Quantum Chromodynamics with all the built-in non-perturbative properties \cite{ADAM2020135928, Adam_2015}. In the present paper, we use the latest form of E-RMF Lagrangian to evaluate the EoS with the recently developed G3 \cite{KUMAR2017197} parameter set, and the results are compared with the familiar NL3 \cite{PhysRevC.55.540}, and IU-FSU \cite{PhysRevC.100.025805} forces. 

The paper is organized as follows: In Sub-Sec. \ref{E-RMF}, the relativistic mean-field formalism is briefly described. The coherent density fluctuation model is detailed in Sub-Sec. \ref{cdfm} and the NS properties are calculated in Sub-Sec \ref{NS}. The results and discussions are given in Sec. \ref{RD}. In this section, the mass, radius, moment of inertia, density profile, and weight function of NS obtained from the  E-RMF equation of states are exemplified and explicitly discussed. The parameters $K^{\rm star}$, $S^{\rm star}$, $L_{\rm sym}^{\rm star}$ and $K_{\rm sym}^{\rm star}$ of NS are illustrated in Sub-Sec. \ref{surf}. The summary and concluding remarks are drawn in Section \ref{conclusion}.  
\begin{table}
\label{table1}
\caption{The nuclear matter properties at saturation for the EoS of NL3 \cite{PhysRevC.55.540}, G3 \cite{KUMAR2017197} and IU-FSU \cite{PhysRevC.100.025805} parameter sets. The NM parameters are in MeV, except $\rho_{0}$ which is in fm$^{-3}$. The references are $[a]$,$[b]$, $[c]$ $\&$ $[d]$ \cite{Zyla_2020}, $[e] $\&$ [f]$ \cite{doi:10.1146/annurev.ns.21.120171.000521}, $[g]$ \cite{GARG201855}, $[h] $\&$ [i]$ \cite{DANIELEWICZ20141}, and $[j]$ \cite{zimmerman2020measuring}.}
\scalebox{1.1}{
\begin{tabular}{cccccccccc}
\hline
\hline
\multicolumn{1}{c}{Parameter}
&\multicolumn{1}{c}{NL3}
&\multicolumn{1}{c}{G3}
&\multicolumn{1}{c}{IU-FSU}
&\multicolumn{1}{c}{Empirical/Expt. Value}\\
\hline
$\rho_{0}$ & 0.148 & 0.148 & 0.154 & 0.148 -- 0.185 $[e]$\\
$E/A$ & -16.29 & -16.02 & -16.39 & -15.00 -- 17.00 $[f]$\\
$K$ & 271.38 & 243.96 & 231.31 &220 -- 260 $[g]$\\
$J_0$ & 37.43 & 31.84 & 32.71 & 30.20 -- 33.70 $[h]$\\
$L_{\rm sym}$ & 120.65 & 49.31 & 49.26 & 35.00 -- 70.00 $[i]$ \\
$K_{\rm sym}$ & 101.34 & -106.07 & 23.28 & -174 -- -31 $[j]$\\
$Q_{\rm sym}$ & 177.90 & 915.47 & 536.46 & -----------\\
\hline
\hline
\end{tabular}}
\end{table}
\begin{table}[H]
\label{table2}
\caption{The fitted coefficients $a_i$, $b_i$
and $b_e$ of the Eq. (\ref{efitting}) for NL3, G3 and IU-FSU forces. The values are scaled by $10^{-8}$ factor i.e. each should be multiplied by a factor of $10^{8}$ to get the exact magnitude of the coefficient.}
\renewcommand{\tabcolsep}{0.4cm}
\renewcommand{\arraystretch}{1.5}
\begin{tabular}{ccccccc}
\hline \hline
& NL3  & G3 & IU-FSU  \\
\hline
be& 0.00017  & 0.00011 & 0.00011  \\
b3 & -0.00054 & -0.000085 & -0.000088 \\ 
b4 &  0.00898 & 0.00048 & 0.00043 \\ 
b5 & -0.08078  & -0.00158 & -0.00091 \\ 
b6 & 0.04609  & 0.00346 & 0.00036 \\ 
b7 & -1.774  & -0.00547 & 0.00241 \\ 
b8 & 4.742  & 0.00648 & -0.00592 \\ 
b9 & -8.896 & -0.00579 & 0.00698 \\ 
b10 & 11.65  & 0.00379 & -0.00498 \\ 
b11 & -10.43 & -0.00174 & 0.00227 \\ 
b12 & 6.0713 & 0.00523 & -0.00064 \\ 
b13 & -2.069 & -0.000092 & 0.000105 \\ 
b14 & 0.3132 & 0.000007 & -0.000007 \\ 
a3 & -0.00088 & -0.000198 & -0.00019 \\ 
a4 & 0.02289 & 0.002919 & 0.002475 \\ 
a5 & -0.02539 & -0.01797 & -0.013555 \\ 
a6 & 1.639 & 0.06395 & 0.04266 \\ 
a7 & -6.864 & -0.1472 & -0.08664 \\ 
a8 & 19.60 & 0.2303 & 0.1198 \\ 
a9 & -39.03 & -0.2502 & -0.1156 \\ 
a10 & 54.27 & 0.1891 & 0.07793 \\ 
a11 & -51.73 & -0.09.762 & -0.03605 \\ 
a12 & 32.25 & 0.03281 & 0.01092 \\ 
a13 & -11.84 & -0.00647347 &-0.00194919 \\ 
a14 & 1.945 & 0.000569 & 0.000155 \\
\hline \hline
\end{tabular}
\end{table}
\section{Theory}
\subsection{Effective field theory relativistic mean field model}
\label{E-RMF}
As mentioned earlier, we used  NL3 \cite{PhysRevC.55.540}, G3 \cite{KUMAR2017197} and IU-FSU \cite{PhysRevC.100.025805} parameter sets of the E-RMF Lagrangian. The NL3 set is the stiffest, and the newly reported G3 parameter set provides the softest EoS. The numerical values of nuclear matter (NM) properties at saturation are listed in Table I. The empirical/experimental data are also given for comparison. The nuclear matter incompressibility $K$, which controls the stiffness/softness of the EoS, are 271.38, 243.96, and 231.31 MeV for NL3, G3, and IU-FSU, respectively. The NM symmetry energies are 37.43, 31.84, and 32.71 MeV for the corresponding parameter sets. These values are within the range set by various experimental observations and theoretical predictions (see Table I).

Motivated by the work of Br\"uckner {\it et al.} \cite{PhysRev.168.1184, bruck68}, using the LDA, the momentum space energy functional is converted to the coordinate space through a generator coordinate `$x$'. The detailed procedure can be found in Refs. \cite{ankit2021,PhysRevC.103.024305}. It is worth mentioning that the Br\"uckner energy density functional \cite{PhysRev.168.1184, bruck68} fails to solve the Coester band problem \cite{PhysRevC.5.1135,jeet21PRC}. Consequently, the peak that appears in the symmetry energy at the magic number for heavier nuclei (like Pb isotopes) does not match the appropriate neutron number \cite{jeet21,jeet21PRC}. The coordinate space E-RMF energy density functional within the LDA for NS matter is defined as \cite{ankit2021}: 
\begin{eqnarray}
\label{efitting}
{\cal E} & = & C_k n^{2/3} + C_e n^{4/9} + \sum_{i=3}^{14} (b_i + a_i \alpha^2) n^{i/3},
\end{eqnarray}
where $C_{k} = 0.3 (\hbar^{2}/2M) (3\pi^{2})^{2/3} [(1+\alpha)^{5/3} + (1-\alpha)^{5/3}]$ is the coefficient of the kinetic energy for protons and neutrons and $C_{e} = b_{e} (1-\alpha)^{5/9}$ is the kinetic energy coefficient for electrons and muons, with $b_{e}$ as a variable obtained from the conversion of the E-RMF energy density from momentum space to coordinate space \cite{ankit2021}. The last term is the potential interaction of the nucleons and the coefficients $b_{i}$ and $a_{i}$  obtained from the fitting for different E-RMF models. It is shown in Ref. \cite{PhysRevC.103.024305} that the accuracy of the fitting increases with increase in the number of coefficients $a_i$ and $b_i$ in the series of the potential  term of Eq. (\ref{efitting}). The mean deviation $\delta=\sum_{j=1}^{N} [(E/A)_{j,\mathrm{Fitted}} - (E/A)_{j,\mathrm{RMF}}]/N$, = $18\%$, $6\%$ and $0.5\%$ for  8, 10 and 12 terms, respectively. Here $N$ is the total number of points. The obtained coefficients of the energy functional in Eq. (\ref{efitting}) i.e. $b_e$, $b_i$ and $a_i$ are tabulated in Table \ref{table1}.
\subsection{Coherent density fluctuation model}
\label{cdfm}
The CDFM is a well-established formalism to calculate the properties of finite nuclei \cite{Antonov1980, PhysRevC.50.164, Gaidarov2020ProtonAN} by superimposing the structure of infinite nuclear matter. This method is recently extended to calculate the properties of NS \cite{ankit2021}. The CDFM utilise a generator coordinate `$x$' to evaluate the one-body density matrix $n(r,r^\prime)$ of a finite nucleus/NS as the superposition of infinite number of one-body density matrices $n_{x}(r,r^\prime)$, called ``Fluctons" \cite{gaid11, JPG47(2020)105102}. The density of a Flucton is written as \cite{Antonov1980, PhysRevC.50.164, Gaidarov2020ProtonAN}:
\begin{equation}
\label{Eq2}
n_x ({\bf r}) = n_0 (x)\, \Theta (x - \vert {\bf r} \vert).
\end{equation}
The saturation density of the Flucton is $n_{0}(x) = 3 A / 4 \pi x^{3}$, $A$ is the total number of protons and neutrons in the neutron star matter (NSM). In the CDFM, the density of the spherical finite NSM of radius `$r$' is \cite{Antonov_2018, Antonov_2016},
\begin{eqnarray}
\label{Eq3}
n (r) &=& \int_0^{\infty} dx\, \vert F(x) \vert^2\, n_{0}(x) \, \Theta(x-\vert{\bf r} \vert),
\label{rhor}
\end{eqnarray}
where $\vert F(x) \vert^2$ is the weight function in the generator coordinate `$x$' with the local density $n(r)$ written as \cite{Antonov_2018}:
\begin{equation}
\label{Eq4}
|F(x)|^2 = - \frac{1}{n_0 (x)} \frac{dn (r)}{dr} \Bigg|_{r=x}.
\end{equation}
The incompressibility, symmetry energy, slope parameter, and curvature coefficient of the NS are expressed by folding the weight function with the respective NS matter as \cite{ankit2021,gaid11, PhysRevC.85.064319, Antonov_2016}:
\begin{eqnarray}
K^{\rm star} &=& \int_0^{\infty} dx\, \vert F(x) \vert^2 \ K^{\rm NSM} (n (x)),
\label{K0} \\
S^{\rm star} &=& \int_0^{\infty} dx\, \vert F(x) \vert^2\, S^{\rm NSM} (n (x)) ,
\label{s0} \\
L_{\rm sym}^{\rm star} &=& \int_0^{\infty} dx\, \vert F(x) \vert^2 \,L_{\rm sym}^{\rm NSM} (n (x)) ,
\label{L0} \\
K_{\rm sym}^{\rm star} &=& \int_0^{\infty} dx\, \vert F(x) \vert^2 \ K_{\rm sym}^{\rm NSM} (n (x)),
\label{c0}
\end{eqnarray}
where $K^{\rm NSM}$, $S^{\rm NSM}$, $L^{\rm NSM}_{\rm sym}$ and  $K^{\rm NSM}_{\rm sym}$ are the incompressibility, symmetry energy, slope parameter and curvature of the NSM.

The converted energy density functional of the NSM from momentum space to the coordinate space `$x$' in a local density approximation is Eq. (\ref{efitting}). The expressions for $K^{\rm NSM}$, $S^{\rm NSM}$, $L^{\rm NSM}_{\rm sym}$ and $K^{\rm NSM}_{\rm sym}$ are obtained from  this Eq. (\ref{efitting}) with the definitions \cite{Fetter, PhysRevC.80.014322, PhysRevC.90.044305}, i.e.,
the NM parameters $K^{\rm NM}$, $S^{\rm NM}$, $L_{\rm sym}^{\rm NM}$ and $K_{\rm sym}^{\rm NM}$ are obtained from the following standard relations \cite{gaid11,Antonov_2016,PhysRevC.85.064319,ankit2021}:
\begin{eqnarray}
K^{\rm NM}&=&9\rho_0^2\frac{\partial^2 ({\cal E}/\rho)}{\partial \rho^2} \Big|_{\rho=\rho_0} \label{knm},\\
S^{\rm NM}&=&\frac{1}{2}\frac{\partial^2 ({\cal E}/\rho)}{\partial\alpha^2}\Big|_{\alpha=0},\label{snm}\\
L_{\rm sym}^{\rm NM}&=&3\rho_0\frac{\partial S(\rho)}{\partial\rho}\Big|_{\rho=\rho_0} = \frac{3P}{\rho_{0}},\label{lsymnm}\\
K_{\rm sym}^{\rm NM}&=&9\rho_0^2\frac{\partial^2 S(\rho)}{\partial\rho^2}\Big|_{\rho=\rho_0},\label{ksymnm}
\end{eqnarray}
which are given as follows using Eq. (\ref{efitting})
\begin{eqnarray}
K^{\rm NSM} &=& -150.12\,n_0^{2/3}(x) - 2.22\,b_{e}\,n_{0}^{4/9}(x) \nonumber \\ 
&&
+ \sum_{i=4}^{14} i\, (i-3)\, b_i\, n_0^{i/3}(x), \label{eqA}\\
S^{\rm NSM} &=& 41.7\,n_0^{2/3}(x) - 0.12\,b_{e}\,n_{0}^{4/9}(x) \nonumber \\
&&
+ \sum_{i=3}^{14} a_i\, n_0^{i/3}(x), \label{eqB}\\
L_{\rm sym}^{\rm NSM} &=& 83.4\,n_0^{2/3}(x) - 0.16\,b_{e}\,n_{0}^{4/9}(x) \nonumber \\
&&
+ \sum_{i=3}^{14} i\, a_i\, n_0^{i/3}(x), \label{eqC}\\
K_{\rm sym}^{\rm NSM} &=& -83.4\,n_0^{2/3}(x) + 0.266\,b_{e}\,n_{0}^{4/9}(x) \nonumber \\
&&
+ \sum_{i=4}^{14} i\, (i-3)\, a_i\, n_0^{i/3}(x). \label{eqD}
\end{eqnarray}
The symmetry energy for finite nuclei, i.e., NS with nucleon number A, can be further expressed as the components of volume $S_V$ and surface $S_S$ contributions using Danielewicz's liquid drop prescription, which is written as \cite{DANI03,DANI04,DANI07,Antonov_2018,Antonov_2016,jeet21,jeetcjp}:
\begin{eqnarray}
S= \frac{S_{V}}{1+ \frac {S_{S}} {S_{V}} A^{-1/3}}= \frac{S_{V}}{1+A^{-1/3}/\kappa},
\label{Eq17}
\end{eqnarray}
where the ratio $\kappa \equiv \frac{S_{V}}{S_{S}}$ is defined as \cite{DANI03,DANI04,DANI07,jeet21,jeetcjp}:
\begin{eqnarray}
\kappa = \frac{3}{R\rho_{0}}\int_0^{\infty} dx \vert F(x) \vert^2 x \rho_{0}(x)
\left [\left (\frac{\rho_{0}}{\rho(x)}\right )^{\gamma}-1\right].
\label{k0t}
\end{eqnarray}
The value of $\gamma=0.3$ is used in Eq. (\ref{k0t}) following Ref. \cite{Antonov_2018}.
An alternative method has been reported by Gaidarov \textit{et al.} to obtain the volume, and surface symmetry energy components \cite{gaid21}.
\subsection{Neutron star properties}
\label{NS}
The neutron star EoS is calculated using the E-RMF model with the assumption that the NS is in $\beta-$equilibrium and charge neutrality conditions \cite{NKGb_1997}. The EoS of the NS depends on the model and also on the types of extra particles such as hyperons \cite{Schaffner_1996, Bipasa_2014, Fortin_2017, KumarTide_2017, Bhuyan_2017, Biswalaip_2019, Biswal_2019}, kaons \cite{Pal_2000, Gupta_2012}, dark matter \cite{Das_2019, 10.1093/mnras/staa1435, JCAP01(2021)007, DasMNRAS_2021, DasPRD_2021} etc. are present in the systems. Here, we limit to the  nucleons and leptons only, which
are in $\beta-$ equilibrium and charge neutrality.
\\
To calculate the NS macroscopic/structural properties such as $M$ and $R$, one has to solve the Tolman-Oppenheimer-Volkoff (TOV) equations \cite{TOV1, TOV2}
\begin{eqnarray}
\frac{dP}{dr}&=&-\frac{[{\cal E}+P] [m+{4\pi r^3P}]}{r^2\Big(1-\frac{2m}{r}\Big)},
\nonumber
\\
\frac{dm}{dr} &=& 4\pi r^2 {{\cal E}}.
\label{eq:TOV}
\end{eqnarray}
The coupled equations are solved by using boundary conditions as: $r=0$, $P=P_c$ and $r=R$, $P=0$ at fixed central density. The maximum mass and radius of the NS are calculated assuming the pressure vanishes at the surface of the star.

For slowly and uniformly rotating NS, the  metric is given by \cite{Stergioulas_2003} 
\begin{eqnarray}
ds^2= -e^{2\nu}dt^2+e^{2\psi}(d\phi - \omega dt^2)+e^{2\alpha}(r^2d\theta^2+ d\phi^2).
\end{eqnarray}
The moment of inertia ($I$) of the NS is calculated with the slow rotation approximation and is  given as \cite{Stergioulas_2003,Jha_2008,Sharma_2009,Friedmanstergioulas_2013,Paschalidis_2017,Quddus_2020,Koliogiannis_2020}:
\begin{equation}
I \approx \frac{8\pi}{3}\int_{0}^{R}\ dr \ ({\cal E}+P)\  e^{-\phi(r)}\Big[1-\frac{2m(r)}{r}\Big]^{-1}\frac{\Bar{\omega}}{\Omega}\ r^4,
\label{eq:moi}
\end{equation}
where $\Bar{\omega}$ is the dragging angular velocity for a uniformly rotating star. The $\Bar{\omega}$ satisfies the following boundary conditions, 
\begin{equation}
\Bar{\omega}(r=R)=1-\frac{2I}{R^3},\qquad \frac{d\Bar{\omega}}{dr}\Big|_{r=0}=0 .
\label{eq:omegabar}
\end{equation}
\section{Results and Discussions}
\label{RD}
In this section, we present the macroscopic properties of the NS, such as mass ($M$), radius ($R$), and moment of inertia ($I$) for NL3, G3, and IU-FSU parameter sets. After getting a broad knowledge of the bulk properties, we extend our calculations to the surface properties of NS with the above three forces. For this, we estimate the symmetry energy $S^{\rm star}$, incompressibility $K^{\rm star}$, slope parameter $L^{\rm star}_{\rm sym}$ and curvature co-efficient $K^{\rm star}_{\rm sym}$ of NS  with respect to their masses from $0.8 \ M_\odot$ to $M_{\rm max}$. To calculate these properties, the NS densities are extracted by feeding the EoSs in the TOV equations. Considering the obtained density as the local density of the star, with the help of CDFM, we construct the weight function $|F(x)|^2$, which is folded with $S^{\rm NSM}$, $K^{\rm NSM}$, $L_{\rm sym}^{\rm NSM}$ and $K_{\rm sym}^{\rm NSM}$, respectively to evaluate the surface properties of the NS. The detailed procedure of the evaluation scheme of the results is available in Ref. \cite{ankit2021}. Further, the results are discussed in the following sub-sections.
\begin{figure}
\centering
\includegraphics[width=0.48\textwidth]{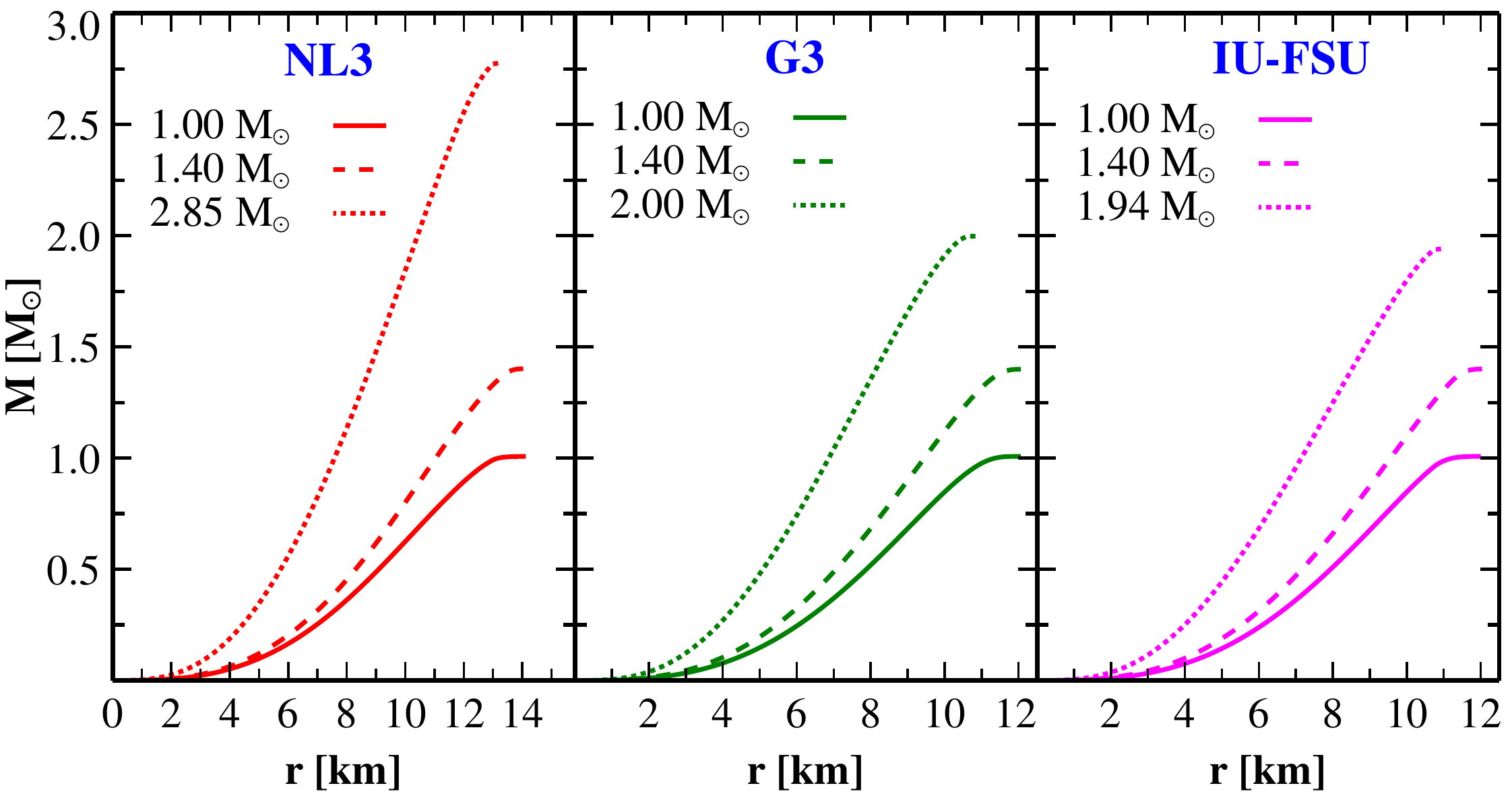}
\caption{(color online) Internal and the maximum mass-radius profiles of a NS for NL3 (red), G3 (green) and IU-FSU (magenta) parameter sets. The different masses $1.0\ M_{\odot}$, $1.4\ M_{\odot}$ and $M_{\rm max}$ are shown with solid, dashed and dotted lines.}
\label{fig:mr}
\end{figure}
\begin{figure}
\centering
\includegraphics[width=0.48\textwidth]{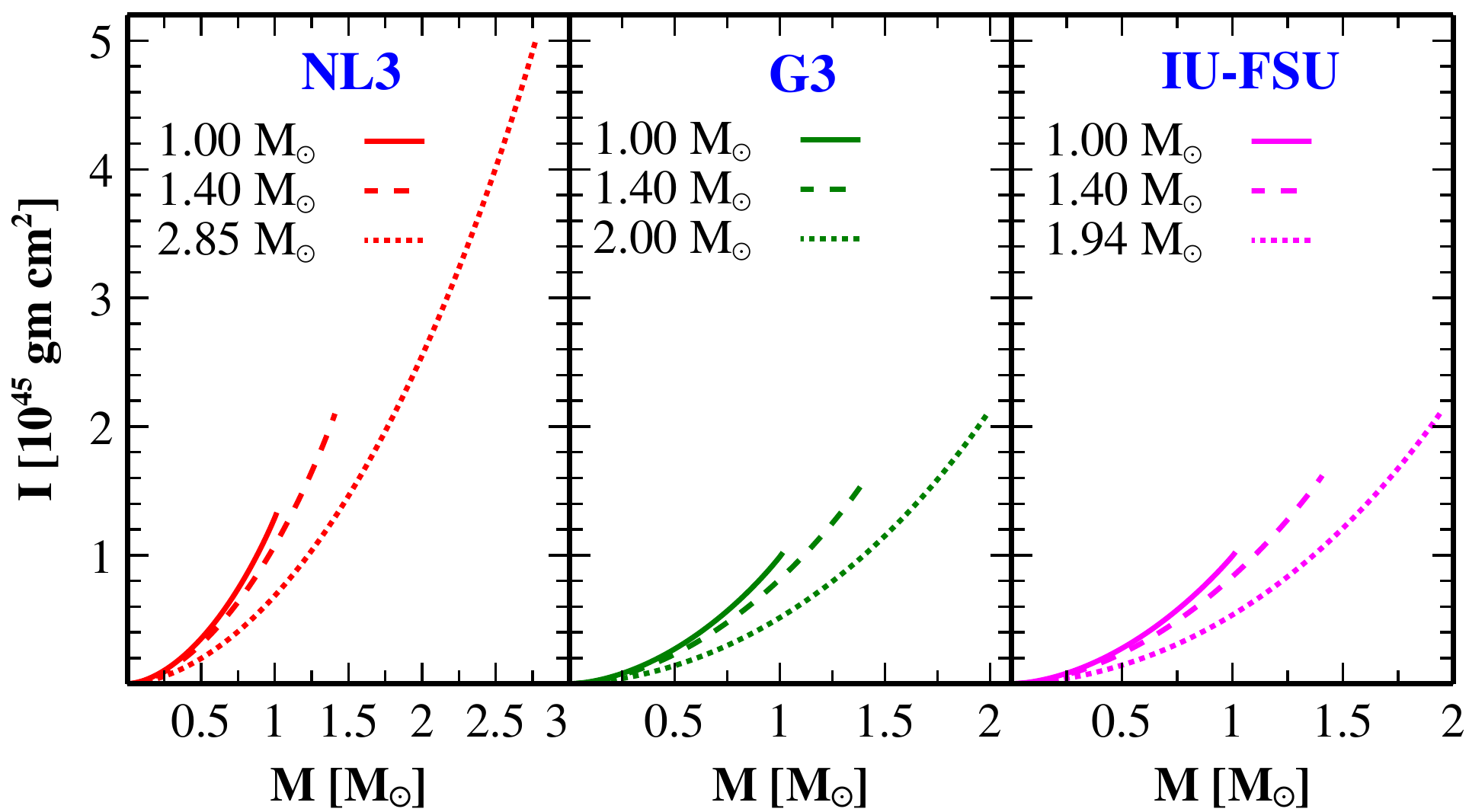}
\caption{(color online) Moment of inertia of NS for NL3 (red), G3 (green) and IU-FSU (magenta) parameter sets. The different masses $1.0\ M_{\odot}$, $1.4\ M_{\odot}$ and $M_{\rm max}$ are shown with solid, dashed and dotted lines.}
\label{fig:mom}
\end{figure}
\subsection{Mass, radius and momentum of inertia of neutron star}
\label{NS_prop}
The mass-radius profiles of the NS are calculated with three different cases $M=1.0\ M_{\odot}$, $1.4\ M_{\odot}$ and $M_{\rm max}$ by fixing the central densities, which are depicted in Fig. \ref{fig:mr} using NL3, G3, and IU-FSU parameter sets. Being the stiffest EoS, NL3 predicts the maximum mass. Since we fix the central densities corresponding to these three masses $1.0\ M_{\odot}, 1.40\ M_\odot$ and $M_{\rm max}$, the radii are also found to be different for each of the parameter sets. This can be seen clearly from Fig. \ref{fig:mr}. The NL3 set predicts both larger mass and radius as compared to G3 and IU-FSU forces. Similarly, we calculate the moment of inertia $I$ of the NS for these three sets, which are shown in Fig. \ref{fig:mom}. The value of $I$ increases with the mass of the NS due to their nearly linear relationship. The NL3 predicts higher $I$ as compared to G3 and IU-FSU. This is because of the stiffer EoS of NL3 than G3 and IU-FSU sets.
\subsection{Neutron star density and it's weight function}
\label{den_wtfn}
\begin{figure}
\centering
\includegraphics[width=0.5\textwidth]{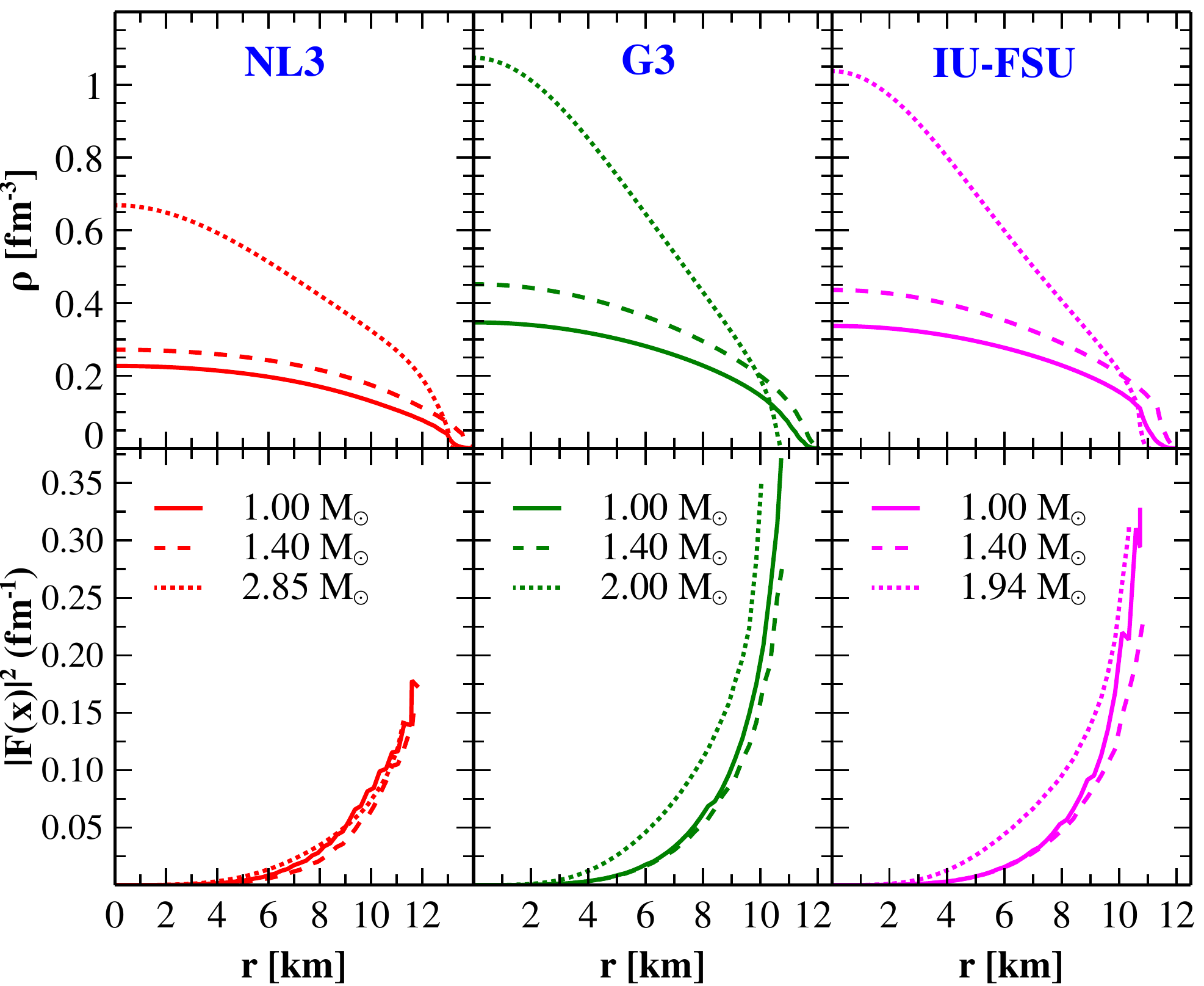}
\caption{(color online) The NS densities ($\rho$) for NL3 (red), G3 (green) and IU-FSU (magenta) parameter sets as a function of radius. The mass number ($A$) of the maximum mass of the NS for NL3, G3 and IU-FSU are $3.35\times10^{54}$, $2.32\times10^{54}$ and $2.23\times10^{54}$ respectively. The different masses $1.0\ M_{\odot}$, $1.4\ M_{\odot}$ and $M_{\rm max}$ are shown with solid, dashed and dotted lines.  }
\label{den}
\end{figure}

The densities and their corresponding weight functions ${|F(x)|^2}$ versus radius of the NS with masses $1.0\ M_{\odot}$, $1.4\ M_{\odot}$ and $M_{\rm max}$ are depicted in Fig. \ref{den}. The densities are in the upper panel, and their weight functions are in the lower panel. The results are presented for NL3, G3, and IU-FSU parameter sets. The chosen masses cover the lower, canonical, and maximum mass of the NS. The maximum mass for NL3, G3, IU-FSU are $2.85\ M_\odot$, $2.004\ M_\odot$, $1.940\ M_\odot$ respectively. The $M_{\rm max}$ for NL3 is distinctly larger than the other two sets. This behavior reflects not only in the $I-M$ and $M-R$ profiles but is also clearly seen in the densities and weight functions. Unlike the normal nucleus, which is bound by strong interaction, the NS is balanced by the attractive gravitational and the repulsive force due to the degenerated neutrons gas. It is worthy of mentioning that the nuclear force is state-dependent, i.e., (i) singlet-singlet, (ii) triplet-triplet, and (iii) singlet-triplet. The former two interactions are attractive, while the latter category is repulsive interaction \cite{EPL20(1992)87, JPG30(2004)771, JPG47(2020)105102, JCAP01(2021)007}. Because of the excessive neutrons in the NS, the repulsive part is subject to instability, which is counterbalanced by the huge gravitational attraction. Thus, NS's density distribution is quite different from the normal nucleus. In addition, the density is influenced by the presence of electrons and muons, and the density obtained by the NL3 set has the minimum central density followed by IU-FSU and G3 models. However, the $M_{\rm max}$ of NL3 is more as compared to the maximum mass of the star acquired by the G3 and IU-FSU parameter sets.

The corresponding weight functions for NL3, G3, and IU-FSU sets are given below to their densities in Fig. \ref{den}. The shape of the  ${|F(x)|^2}$ is like an exponential rise, and it is maximum at the surface of the NS. The values of $K^{\rm star}$, $S^{\rm star}$, $L^{\rm star}_{\rm sym}$ and $K_{\rm sym}^{\rm star}$ are determined by folding  the weight function with the corresponding NSM quantities are  $K^{\rm NSM}$, $S^{\rm NSM}$, $L_{\rm sym}^{\rm NSM}$ and $K_{\rm sym}^{\rm NSM}$ (see Eqs. \ref{K0}, \ref{s0}, \ref{L0}, \ref{c0}). Thus, the maximum contribution comes from the surface of the NS and is termed a surface phenomenon. Precisely, the values of ${|F(x)|^2}$ gather momentum at $\sim{6}$ km, and it is maximum at the surface ($\sim{10-12}$ km).
\begin{figure}
\centering
\includegraphics[width=0.4\textwidth]{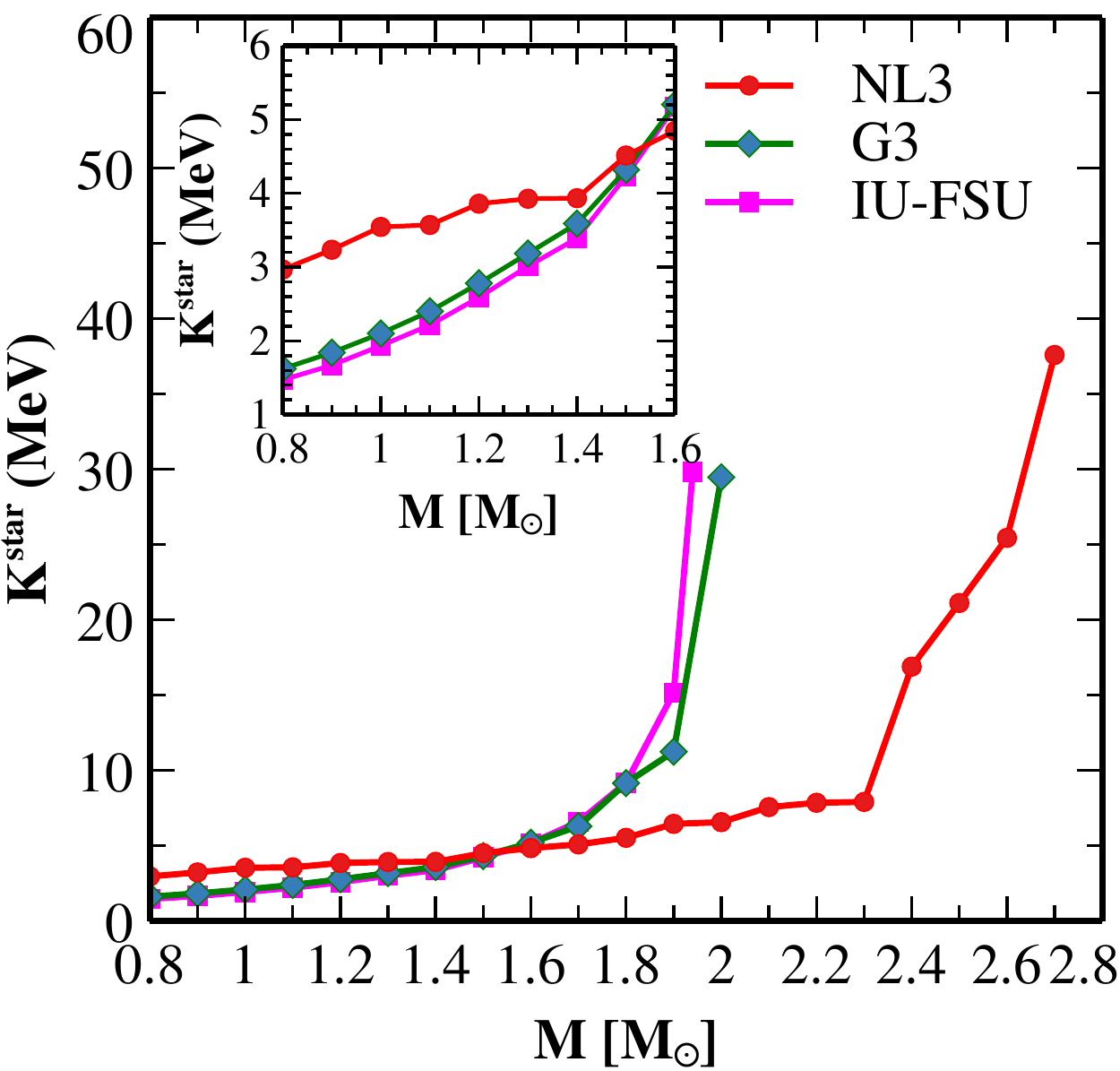}
\caption{(color online) The NS incompressibility $K^{\rm star}$ 
as a function of NS mass for NL3 (red), G3 (green) and IU-FSU (magenta) parameter sets.
\label{fig4}
}
\end{figure}
\begin{figure}
\centering
\includegraphics[width=0.4\textwidth]{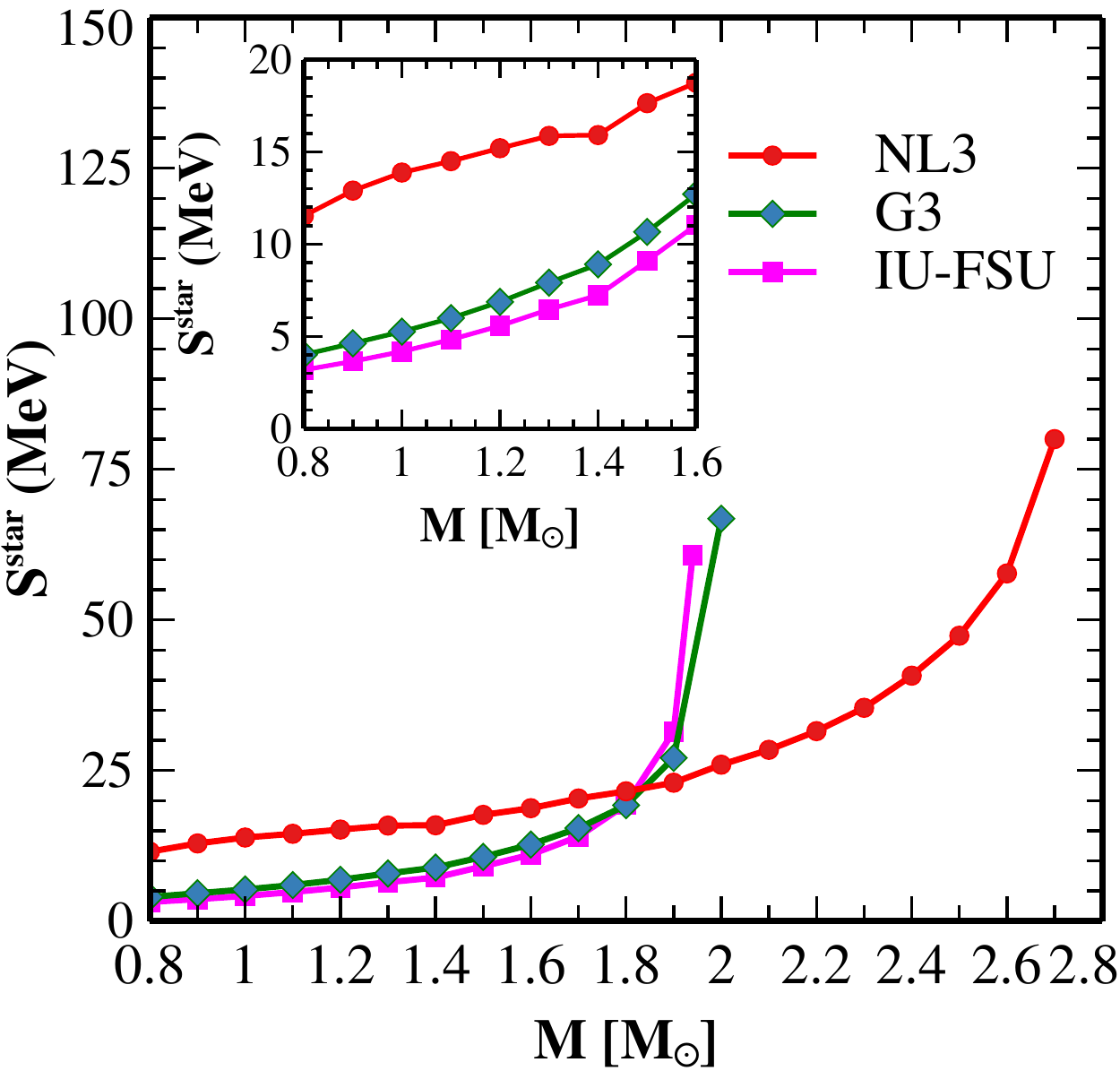}
\caption{(color online) The symmetry energy of the NS $S^{\rm star}$ as a function of NS mass for NL3 (red), G3 (green) and IU-FSU (magenta) parameter sets.}
\label{fig5}
\end{figure}
\begin{figure}
\centering
\includegraphics[width=0.4\textwidth]{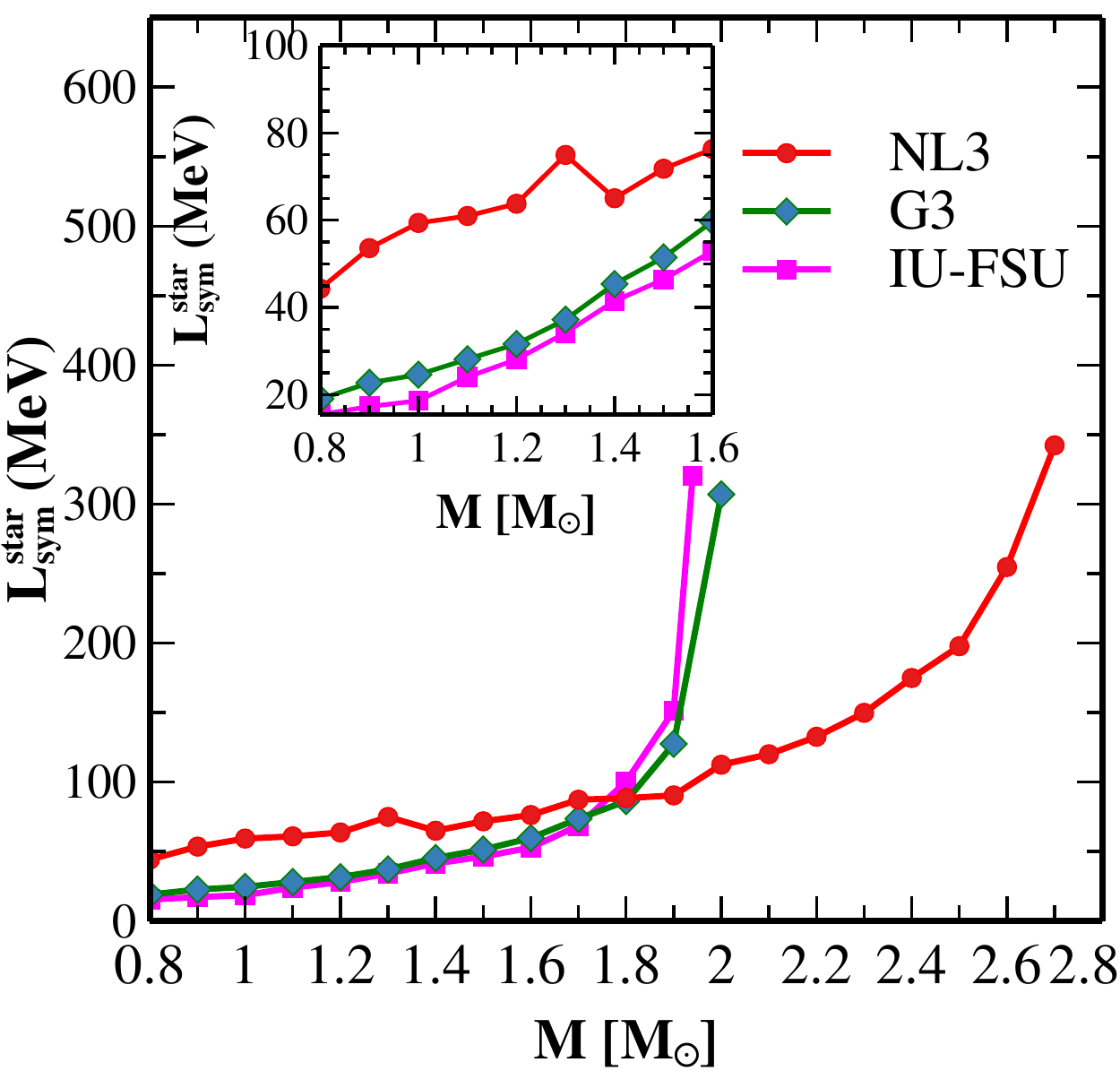}
\caption{(color online) The $L^{\rm star}_{\rm sym}$ coefficient of the NS $S^{\rm star}$ obtained from Eq. (\ref{L0}) as a function of NS mass for NL3 (red), G3 (green) and IU-FSU (magenta) parameter sets.}
\label{fig6}
\end{figure}
\begin{figure}
\centering
\includegraphics[width=0.4\textwidth]{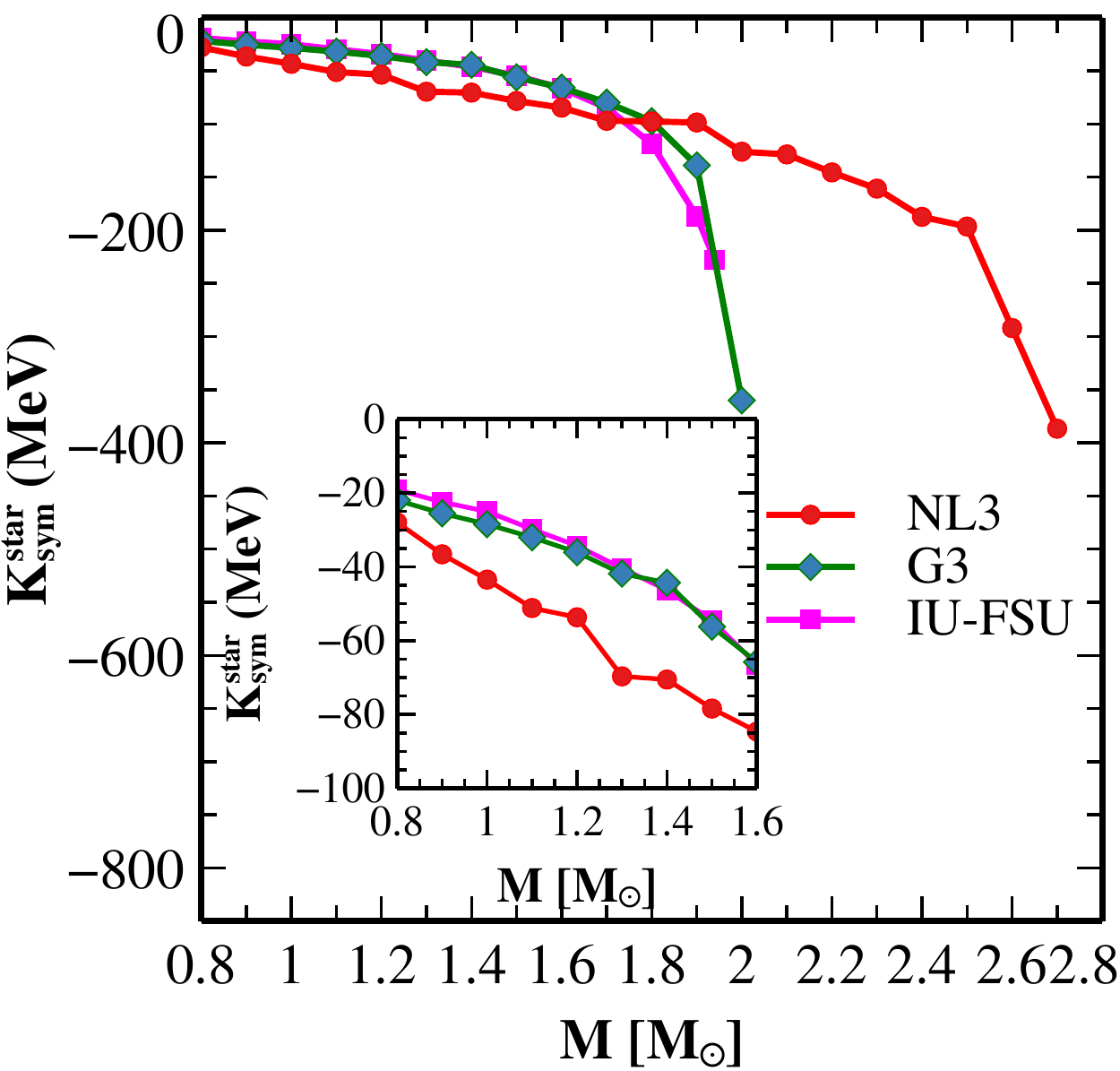}
\caption{(color online) 
The curvature coefficient $K^{\rm star}_{\rm sym}$  of the NS obtained from Eq. (\ref{c0}) as a function of NS mass for NL3 (red), G3 (green) and IU-FSU (magenta) parameter sets.}
\label{fig7}
\end{figure}
\begin{figure}
\centering
\includegraphics[width=0.4\textwidth]{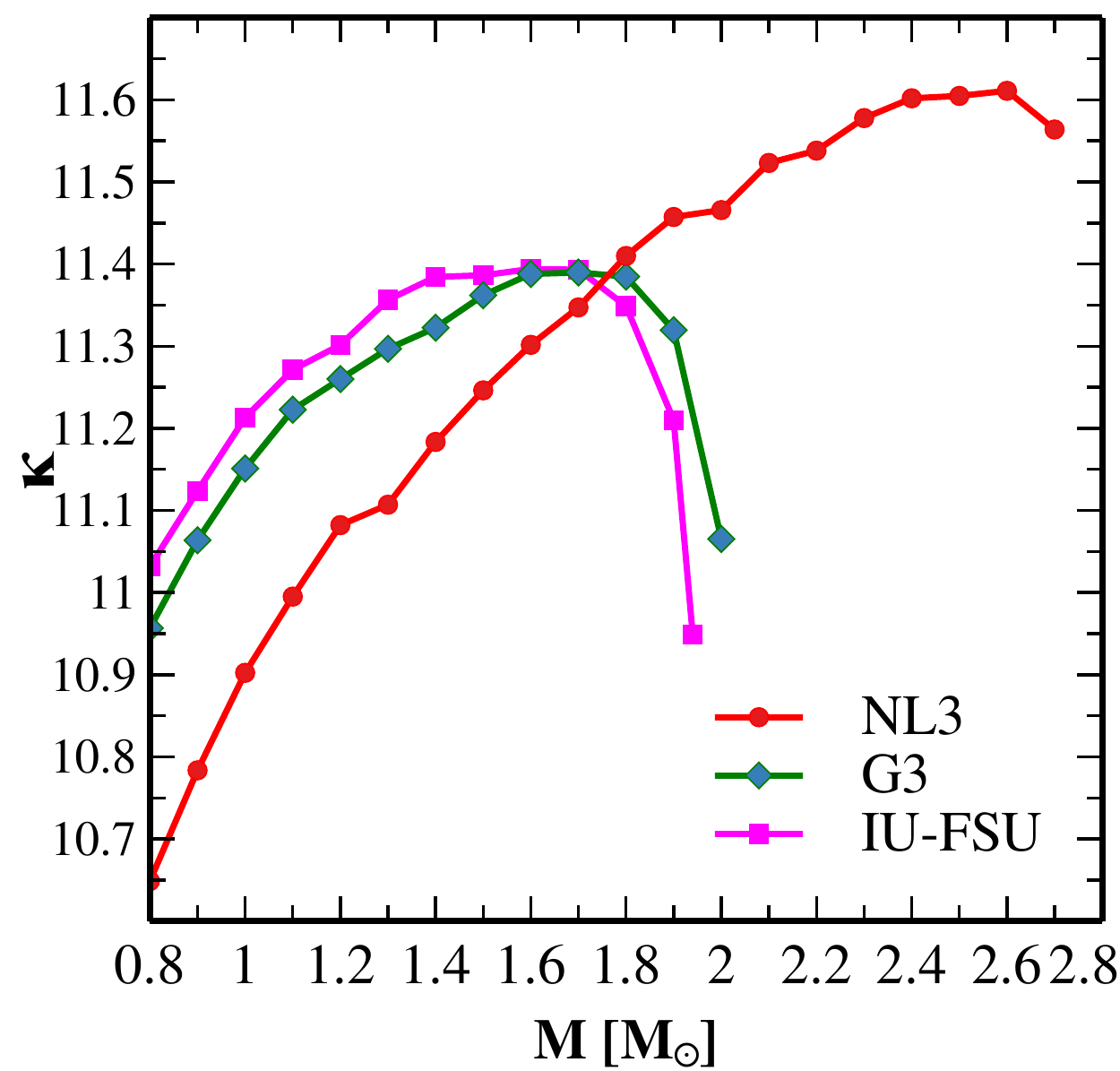}
\caption{(color online) The ratio of the volume to surface components of the symmetry energy $\kappa$ as a function of NS mass for NL3 (red), G3 (green) and IU-FSU (magenta) parameter sets. The $\kappa$ values are evaluated from Eq. (\ref{k0t}).}
\label{fig8}
\end{figure}
\subsection{Surface properties of neutron star}
\label{surf}
In this sub-section, we analyse the surface properties of our results obtained from the  CDFM calculations for NS as a function of mass. Here, the incompressibility $K^{\rm star}$, symmetry energy $S^{\rm star}$, slope parameter $L^{\rm star}_{\rm sym}$ and curvature
co-efficient $K^{\rm star}_{\rm sym}$ of the NS are discussed. The results are depicted in Figs. (\ref{fig4},\ref{fig5},\ref{fig6},\ref{fig7}) in the following sub-sections.
\subsubsection{Neutron star incompressibility}
The incompressibility $K$ of an object is a prominent characteristic to know its nature. It defines how much the object can be compressed or expanded. So, it has a direct connection with the collective motion of the system. As a matter of fact, $K$ is a significant quantity for the NS. The EoS, governed by incompressibility, plays a crucial role in determining the mass and radius of the NS. It is shown in Ref. \cite{ankit2021} that the incompressibility of NS  $K^{\rm star}$is much less than the nuclear matter incompressibility at saturation $K_{\infty}$. For example, $K_{\infty}=271.38$ MeV for NL3 set as compared to the  $K^{\rm star}=44.956$ MeV for the maximum mass of the NS \cite{ankit2021}. This can also be related to the asymmetric nature of the medium $\alpha$. The $K_{\infty}$ is obtained at the asymmetric limit $\alpha=0$, and $K^{\rm star}$ is evaluated at $\alpha\sim 1$, i.e., the incompressibility of a  system decreases with asymmetric of the system \cite{ankit2021}. One can see the trend of $K^{\rm star}$ as a function of mass $M$ ranging from $M=0.8\ M_{\odot}-M_{max}$ of the NS in Fig. \ref{fig4} for the three considered sets. The three forces predict almost similar incompressibilities up to mass $M=1.8\ M_\odot$. More explicitly, the G3 and IU-FSU give almost similar results, while NL3 predicts a comparatively larger value of $K^{\rm star}$ as can be seen from the inset of the figure. Beyond mass $M=1.8\ M_\odot$, the incompressibility increases suddenly. The similarity of $K^{\rm star}$ between G3 and IU-FSU forces could be related to the $K_{\infty}$, which are 243.96 and 231.31 MeV for G3 and IU-FSU, respectively. On the other hand, the incompressibility of NL3 at saturation is quite high as compared to G3 and IU-FSU. After realizing that the EoS can be made softer by reducing the $K$, which in turn reduces considerably the maximum mass of the NS and vice-versa. This shows that the value of $K^{\rm star}$ increases marginally up to $M=1.8\ M_{\odot}$ and suddenly increases beyond, indicating the mass dependence of the incompressibility.
\subsubsection{Symmetry energy and its higher-order derivatives}
The symmetry energy of the NS in its maximum mass is predicted to be higher in comparison to the value of symmetric nuclear matter at saturation. This observation is noticed in all the three considered E-RMF models. The symmetry energy at saturation $J_0$ for NM with NL3, G3, and IU-FSU sets are $37.43$, $31.84$, and $32.71$ MeV, respectively. These results are $146.002$, $66.813$, and $60.758$ MeV for the NS at the limit of maximum mass. These $S^{\rm star}$ are small for smaller masses as compared to the maximum mass of the star. The results of $S^{\rm star}$ are depicted in Fig. \ref{fig5} as a function of mass for all the three-parameter sets of NS. The symmetry energy for G3 and IU-FSU are found to be almost similar, while the values with the NL3 set are a bit higher. This can be seen clearly from the inset of the figure, which depicts the structural dependence of the symmetry energy. The symmetry energy is obtained from the derivative of the energy density with respect to asymmetricity $\alpha$, which shows a significant variation in the $S^{\rm star}$ as compared to $K^{\rm star}$. The higher derivatives of the symmetry energy, i.e., the slope $L^{\rm star}_{\rm sym}$ and curvature parameter $K^{\rm star}_{\rm sym}$ are quite useful quantities. These are shown in Figs. \ref{fig6} and \ref{fig7} as a function of NS mass. The magnitude of $L^{\rm star}_{\rm sym}$ and $K^{\rm star}_{\rm sym}$ increases with the mass of the star. The $L^{\rm star}_{\rm sym}$ values for all the models are found to be positive, contrary to the negative nature of $K^{\rm star}_{\rm sym}$ as shown in the figures.

The negative sign of $K^{\rm star}_{\rm sym}$ is correlated by the 1-$\sigma$ constraint, and $90\%$ confident limits on its saturation value of normal nuclear matter as reported by Zimmerman et al. from the experimental data of PSR J0030+0451 and GW170817 event \cite{zimmerman2020measuring, Riley_2019, PhysRevLett.119.161101}. It is noted that these bounds are not well matched to explain the $K^{\rm star}_{\rm sym}$ of NS; however, it indicates the possibility of a negative value of the curvature parameter and predicts the range with $\sim{90}\%$ confidence limit of the observational data. Sometimes it is beneficial to analyze the terrestrial data related to the exotic nuclei and heavy-ion collisions \cite{PhysRevC.76.054316, PhysRevLett.102.122502} by separating the contribution of iso-vector incompressibility or curvature parameter.

Finally, the total symmetry energy of the NS can be divided into its volume $S_V$, and surface $S_S$ components, which are derived from Eq. (\ref{Eq17}) through $\kappa$ (the ratio of the volume to surface symmetry energies) and $\kappa$ is obtained from Eq. (\ref{k0t}). The value of $\kappa$ as a function of the mass of the NS is shown in Fig. \ref{fig8}. In our calculation, $\gamma=0.3$ is used following Ref. \cite{Antonov_2018}. The $\kappa$ values are consistently larger for the NL3 set followed by IU-FSU and least for the G3 force. These trends are in accordance with the forces used in the calculations. It goes on increasing with the mass of the NS to some value, as shown in the figure. Beyond that, the $\kappa$  decreases considerably. When the mass number of the system approaches a large value, i.e., in the limit of $A{\rightarrow{\infty}}$, the system deals mostly with the volume. For example, for 1.4 $M_\odot$ of the G3 set, the total symmetry energy is 8.904 MeV, while the volume part contributes as a whole and the surface component contributes nearly 9\% only of total $S^{\rm star}$. In such a case, the major contribution of symmetry energy comes from the volume part of the NS as referring Eq. (\ref{Eq17}). To take care of such a system properly, the alternative method of Ref. \cite{gaid21} may be useful. The $\kappa$ is defined as $\kappa=S_V/S_S$, and even though the total symmetry energy is contributed by the volume component, i.e., $S\to S_V$, it has a surface component due to the finite value of $\kappa$.

\section{Summary and conclusions}
\label{conclusion}
The structural properties of NS, along with the mass, radius, and moment of inertia, are studied within the well-known E-RMF formalism. Three established forces (NL3, G3, and IU-FSU) are used in the calculations. Although NL3 is one of the oldest sets, it gives an understanding of the properties of finite nuclei. Also, this set is used extensively so that it may provide a piece of known information, and a comparison of other sets with the results of NL3 may be more familiar. The primary motivation of the present work is the surface properties of NS in terms of incompressibility, symmetry energy, and its higher coefficients, such as the slope and curvature coefficients. 

To our knowledge, there are no prior theoretical results or empirical/experimental data are available to support our calculations of NS symmetry energy, incompressibility, slope, and curvature parameters. In the recent work, we suggested the formalism for the computation of these quantities \cite{ankit2021} for NS. In the present paper, we extended the model to study the properties systematically. Thus, the results reported here are the first of such kind. We know that the NS is a highly iso-spin asymmetric system. We expected that the NS surface properties could be very different from the standard NM. The more significant $L_{\rm sym}^{\rm star}$ for NL3 force is in agreement with the stiff EoS \cite{sym12060898} and moderate slope parameter predicted for the softer G3 and IU-FSU sets. A proper understanding of the range of $S^{\rm star}$, $L^{\rm star}_{\rm sym}$ and $K^{\rm star}$ generally helps to fix the radius of the NS. A precise correlation is established via Danielewicz's liquid drop prescription with the factor $\kappa$, i.e., the volume ratio to the surface component of the symmetry energy. This correlation can be extended to $L_{\rm sym}^{\rm star}$ in the liquid drop mass formula with the help of surface $S^{\rm star}_S$, and the volume $S^{\rm star}_V$ symmetry energy \cite{2015AIPC.1645...61L}. From the analysis of the surface properties, we noticed that almost all these parameters, along with $\kappa$ for all the three sets, coincide with each other in the vicinity of mass range $M\sim 1.8M_{\odot}$ indicating a possible correlation with the mass of the NS. It is shown in \cite{2015AIPC.1645...61L, Lattimer_2013} that the static dipole and quadrupole polarizability and the neutron-skin thickness are strongly related between symmetry energy and slope parameter. Thus, we expect the NS radius to be synchronized with its surface properties.
The magnitude of $K^{\rm star}$, $S^{\rm star}$, $L^{\rm star}_{\rm sym}$ and $K_{\rm sym}^{\rm star}$ increases with mass of the NS. For smaller NS, we find smaller values of all the surface properties. The second derivative of symmetry energy is the $K_{\rm sym}^{\rm star}$, which is obviously more sensitive to the mass and also most ambiguous.

The present systematic calculations may provide a better theoretical bound on $K^{\rm star}_{\rm sym}$ and a better pathway to constraint the experimental setup for the isoscalar giant resonances for the properties of astrophysical objects. Despite the absence of direct experimental data for these surface properties of NS (incompressibility, symmetry energy, slope, and curvature parameters), the calculated results using the  E-RMF densities and CDFM approach seem reasonable. The present theoretical calculations can be validated using different relativistic and non-relativistic energy density functionals and suitable force parameters. The current method of accessibility to NS, considering the NS as a finite nucleus system, favors a bridge between the two unequal size objects. The theoretical approach adopted here presents a new way for the nuclear and astrophysicist to reveal the wealth of information on exotic nuclei and dense astronomical objects. 

\noindent 
{\bf  Acknowledgement:}  Mr. J. A. Pattnaik thanks the Institute of Physics, Bhubaneswar, for its facilities. SERB partly reinforces this work, Department of Science and Technology, Govt. of India, Project No. CRG/2019/002691.
\bibliography{cdfm}
\end{document}